\documentclass[prl,twocolumn,showpacs]{revtex4-1}
\usepackage[utf8x]{inputenc}
\usepackage{amsbsy}
\usepackage{graphicx}
\usepackage{color}
\usepackage{dcolumn}
\usepackage{amsmath}
\usepackage{amssymb}
\usepackage{amsfonts}
\usepackage{subfigure}

\newcommand{\abinitio}{\emph{ab initio}}
\newcommand{\etal}{\emph{et al.}}

\begin{document}
\title{Anharmonic stabilization of the high-pressure simple cubic phase of calcium}

\author{Ion Errea$^{1, 2, 3}$}
\author{Bruno Rousseau$^{2, 3}$}
\author{Aitor Bergara$^{1, 2, 3}$}

\affiliation{$^1$Materia Kondentsatuaren Fisika Saila, Zientzia eta
Teknologia Fakultatea, Euskal Herriko Unibertsitatea, 644
Postakutxatila, 48080 Bilbao, Basque Country, Spain}
\affiliation{$^2$Donostia International Physics Center (DIPC), Paseo
de Manuel Lardizabal 4, 20018, Donostia, Basque Country, Spain}
\affiliation{$^3$Centro de Física de Materiales CFM - Materials Physics Center MPC, Centro Mixto CSIC-UPV/EHU,
Edificio Korta,
Avenida de Tolosa 72, 20018 Donostia, Basque Country, Spain}

\date{\today}

\begin{abstract}
The phonon spectrum of the high-pressure simple cubic phase of calcium, in the harmonic approximation,
 shows imaginary branches that make it mechanically unstable. In this letter, the phonon spectrum is 
recalculated using density-functional theory (DFT) \abinitio{} methods fully including anharmonic 
effects up to fourth order at 50 GPa. Considering that perturbation theory cannot be employed
with imaginary harmonic frequencies, a variational procedure based on the Gibbs-Bogoliubov
inequality is used to estimate the renormalized phonon frequencies. The results show that
strong quantum anharmonic effects make
the imaginary phonons become positive even at zero temperature
so that the simple cubic phase becomes mechanically stable, as experiments suggest. 
Moreover, our calculations find a superconducting $T_c$ in agreement with experiments 
and predict an anomalous behavior of the specific heat. 
\end{abstract}

\pacs{63.20.kg,63.20.dk,63.20.Ry,74.25.Kc}

\maketitle

The understanding of crystal lattice vibrations in terms of phonons provides an excellent 
paradigm to interpret and understand a wide range of physical phenomena~\cite{born-huang}. 
In most cases, the harmonic approximation describes accurately phonon frequencies and the associated 
physical properties. However, there are examples where experimentally confirmed 
structures display imaginary phonons in \abinitio{} DFT calculations, indicating that 
in such cases anharmonicity cannot be neglected.
The high-pressure simple cubic (sc) phase of calcium is an important 
example of the possible stabilizing role of anharmonicity. Indeed, 
while measurements confirm the presence and stability of this
structure~\cite{Olijnyk1984191,JPSJ.74.2391,PhysRevB.79.134121,Mao01062010}, 
theoretical calculations based on the harmonic approximation find imaginary phonon 
branches throughout the whole Brillouin zone (BZ)~\cite{errea:443,Gao2008181,PhysRevB.78.140101,PhysRevLett.105.235503}.

Under pressure, calcium exhibits a complex and interesting behavior. For instance, it becomes the
element with the largest superconducting critical temperature ($T_c$), 
reaching 25 K at 161 GPa~\cite{JPSJ.75.083703}.
According to room temperature x-ray diffraction 
measurements~\cite{Olijnyk1984191,JPSJ.74.2391,PhysRevLett.101.095503,PhysRevB.81.140106}, 
the ambient condition face-centered-cubic (fcc) phase 
transforms to body-centered-cubic (bcc) at 19 GPa, to sc at 32 GPa, to $P4_32_12$ at 119 GPa, to 
$Cmca$ at 143 GPa and to $Pnma$ at 158 GPa. Moreover, it has recently been reported
that upon cooling the sc structure transforms into a very similar monoclinic $Cmmm$ phase 
at 30 K and 45 GPa~\cite{Mao01062010}.
On the other hand, evolutionary DFT simulations within the generalized
gradient approximation (GGA) at 0 K~\cite{Oganov27042010} found that the experimental phases
do not always coincide with the lowest enthalpy structures. This is quite dramatic in the stability 
range of the sc phase considering that the $I4_1/amd$ structures (from 33 to 71 GPa) 
and $C2/c$ structures (from 71 to 89 GPa) have considerably lower enthalpy than sc.  
Recent diffusion quantum Monte Carlo calculations (DMC)~\cite{PhysRevLett.105.235503}
have brought new insight to this problem, showing that the sc phase is energetically preferred over 
the $I4_1/amd$ phase when the exchange-correlation energy is treated correctly. 
Nevertheless, the question of dynamical stability remains
and a proper quantum-mechanical treatment
explicitly incorporating anharmonicity is still missing. 

The extreme anharmonicity in sc Ca requires a non-perturbative approach and suggests the 
application of the 
self-consistent harmonic approximation (SCHA)~\cite{hooton422,PhysRevLett.17.89}.
The SCHA seeks the physically well-defined Gibbs-Bogoliubov bound and, in contrast to 
classical molecular dynamics (MD) or statistical sampling methods~\cite{PhysRevLett.100.095901},
works at any temperature with no additional cost.  
However, in order to apply this theory,
the knowledge of all anharmonic coefficients is needed. 
Calculating them from first principles is usually complicated and highly time-demanding,
thus, the SCHA has been normally applied making use of empirical potentials.  
Nevertheless, given the simplicity and high-symmetry of the
sc structure, we have calculated \abinitio{} all the necessary
anharmonic coefficients up to fourth order in displacement.
The SCHA could then be applied to compute the
temperature dependent renormalized phonon frequencies.
The calculations have been performed  
at 50 GPa and, as it turns out, within this formalism the phonons of sc 
Ca are stable even at 0 K at this pressure. Unless stated otherwise, we use atomic units throughout,
i.e., $\hbar=1$.

Within the Born-Oppenheimer approximation, the Hamiltonian describing the dynamics of the $N$ ions 
in the crystal is given by $\hat{H} = \hat{T} + \hat{U}$,
where $\hat{T}$ and $\hat{U}$ are, respectively, the kinetic and potential energy operators of the ions.
The potential is expanded up to fourth order as $\hat{U} = \hat{U}_0 + \hat{U}_2 + \hat{U}_3 + \hat{U}_4$ with 
\begin{equation}
\hat{U}_n = \frac{N^{1-\frac{n}{2}}}{n!} \sum_{\{\alpha \mathbf{q}\}} \hat{u}^{\alpha_1}(\mathbf{q}_1) \dots 
\hat{u}^{\alpha_n}(\mathbf{q}_n)
\Phi^{\alpha_1 \dots \alpha_n}(\mathbf{q}_1, \dots, \mathbf{q}_n).
\label{potential}
\end{equation}
In Eq. \eqref{potential}, $\{\alpha\}$ represent Cartesian coordinates, 
$\Phi^{\alpha_1 \dots \alpha_n}(\mathbf{q}_1, \dots, \mathbf{q}_n)$ is the Fourier transform of 
the $n$th derivative of the total energy with respect to the ionic displacements and 
$\hat{u}^{\alpha}(\mathbf{q})$ is the Fourier transform of the ionic displacement operator. 
In the harmonic approximation, neglecting the third and
 fourth order terms of the potential, the Hamiltonian is diagonalized in terms of phonons. 
The term $\hat{U}_3 + \hat{U}_4$ can be treated within perturbation theory to correct the phonon frequencies
and account for their finite lifetime~\cite{PhysRev.128.2589}. However, 
in sc Ca the energy has no lower bound due to the imaginary frequencies obtained in the 
harmonic approximation and, therefore, one needs to treat anharmonicity beyond perturbation theory. 
In the SCHA, one adds and subtracts to the potential a trial 
harmonic term that yields well defined real phonon frequencies, 
$\hat{U}_2^0$, and redefines the Hamiltonian as $\hat{H}=\hat{H}_0+\hat{H}_1$, 
with $\hat{H}_0=\hat{T}+\hat{U}_2^0$ and $\hat{H}_1=(\hat{U}_2-\hat{U}_2^0)+\hat{U}_3 + \hat{U}_4$. 
The exact free energy $F$ satisfies the Gibbs-Bogoliubov inequality
\begin{equation}
F \leq F_0 + \langle \hat{H}_1 \rangle_0,
\label{gibbs-bogoliubov}
\end{equation}
so that the minimum of the right-hand side of Eq. \eqref{gibbs-bogoliubov} becomes an excellent approximation of $F$.
$F_0$ and $\langle \hat{H}_1 \rangle_0$ are given as $F_0 = - \frac{1}{\beta} \ln Z$ and 
$\langle \hat{H}_1 \rangle_0 = \mathrm{tr} ( \hat{H}_1 e^{-\beta \hat{H}_0})/Z$,
where $\beta=1/k_BT$ and the partition function is $Z=\mathrm{tr}( e^{-\beta \hat{H}_0})$.

The adjustable parameters that can be used for the minimization are the trial
 phonon frequencies $\{\Omega_{\nu \mathbf{q}}\}$ that diagonalize $\hat{H}_0$, $\nu$ being a mode index.
Differentiating Eq. \eqref{gibbs-bogoliubov} with respect to $\Omega_{\nu \mathbf{q}}$, the
equation for the trial frequencies that minimize the free energy can be obtained straightforwardly: 
\begin{equation}
\Omega^2_{\nu \mathbf{q}} = \omega^2_{\nu \mathbf{q}} + 8 \Omega_{\nu \mathbf{q}} j_{\nu \mathbf{q}}.
\label{omega-eq}
\end{equation}
A numerical solution of this equation leads to
the renormalized frequencies $\Omega_{\nu \mathbf{q}}$ at any temperature. In Eq. \eqref{omega-eq}
\begin{eqnarray} 
j_{\nu \mathbf{q}} & = & \frac{1}{8N}\sum_{\nu' \mathbf{q}' \{\alpha\}} 
\frac{\epsilon^{\alpha_1}_{\nu' \mathbf{q}'}
\epsilon^{\alpha_2}_{\nu' -\mathbf{q}'}
\epsilon^{\alpha_3}_{\nu \mathbf{q}}
\epsilon^{\alpha_4}_{\nu -\mathbf{q}}
}{4M^2\Omega_{\nu \mathbf{q}}
\Omega_{\nu' \mathbf{q}'}} \nonumber \\ & \times &
\Phi^{\alpha_1 \alpha_2 \alpha_3 \alpha_4}(\mathbf{q}',-\mathbf{q}',\mathbf{q},-\mathbf{q})
[1 + 2 n_B({\Omega_{\nu' \mathbf{q}'}})], \label{gnq}
\end{eqnarray}
$\boldsymbol{\epsilon}_{\nu \mathbf{q}}$ is the phonon polarization vector, $M$ the mass of Ca,
$\{\omega_{\nu \mathbf{q}}\}$ the phonon frequencies 
diagonalizing $\hat{U}_2$, imaginary at some $\bf q$, and $n_B$ the usual bosonic occupation factor.
As it can be seen, the third order anharmonic coefficients do not contribute to $F$ at this level of 
approximation. It should be noted that in the renormalization process the polarization vectors are kept
fixed. This is justified for the highly-symmetric sc phase, but in systems with different atoms in the unit cell
polarization vectors may be used to minimize Eq. \eqref{gibbs-bogoliubov}.

\begin{figure}[t]
\includegraphics[width=0.49\textwidth]{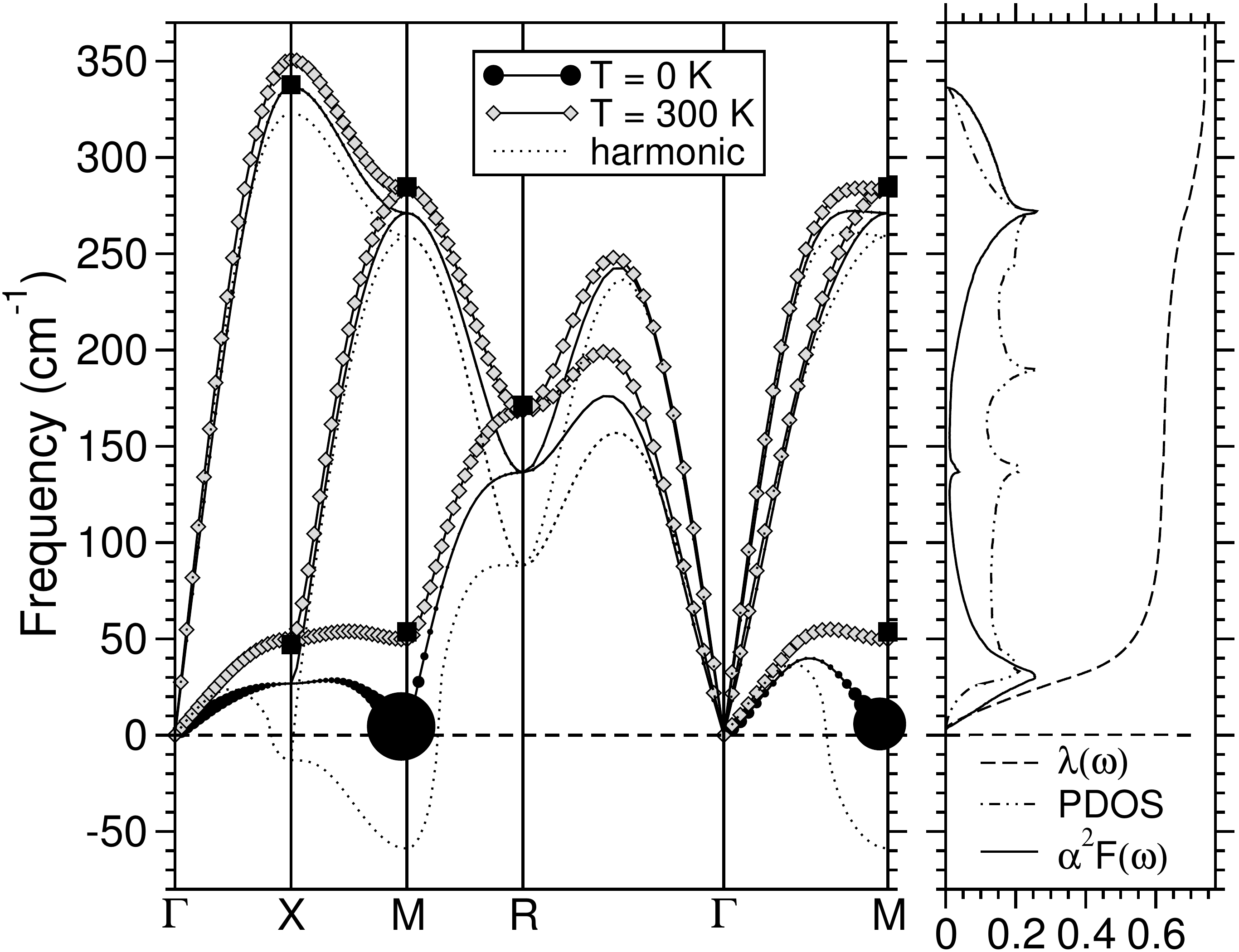}
\caption{(Left panel) Harmonic phonon spectra and renormalized anharmonic phonon
spectra at 0 K and 300 K of sc Ca at 50 GPa. For the 0 K anharmonic branches the value of the mode electron-phonon
coupling, $\lambda_{\nu \mathbf{q}}$, is proportional to the area of each filled circle. Filled
squares depict the renormalized 
frequencies obtained by Teweldeberham \etal{}~\cite{PhysRevLett.105.235503} from classical MD at 300 K and 45 GPa.
(Right panel) At zero temperature, the anharmonic results for the integrated electron-phonon 
coupling parameter $\lambda(\omega)$, the Eliashberg function
$\alpha^2F(\omega)$ and the phonon density of states (PDOS).}
\label{fig1}
\end{figure}

The computationally most expensive part of the method described above is the \abinitio{} calculation of the 
fourth order anharmonic coefficients
 $\{\Phi^{\alpha_1 \alpha_2 \alpha_3 \alpha_4}(\mathbf{q}',-\mathbf{q}',\mathbf{q},-\mathbf{q})\}$. 
These can be obtained taking numerical second derivatives of dynamical matrices calculated in supercells
(see, for example, Ref.~\cite{PhysRevB.82.104504}: 
the method presented there was slightly extended, 
given that two linearly independent real displacements 
must be used to generate the necessary supercells at $\mathbf{q}$ points not at the BZ edge).
Such dynamical matrices were obtained using density functional perturbation theory (DFPT)~\cite{RevModPhys.73.515}
as implemented in {\sc Quantum ESPRESSO}~\cite{0953-8984-21-39-395502} within the PBE-GGA~\cite{PhysRevLett.77.3865} 
and making use of a 10 electron 
ultrasoft pseudopotential with $3s$, $3p$ and $4s$ states in the valence. 
A 30 Ry cutoff was used for the plane-wave basis and a
$16\times16\times16$ Monkhorst-Pack $\mathbf{k}$ mesh for the BZ integrations.
Phonon frequencies and anharmonic coefficients
were calculated on a $4\times4\times4$ $\mathbf{q}$ grid 
\footnote{
The anharmonic coefficients at $\mathbf{q}=2\pi/a[0.25,0.25,0.25]$ 
and symmetry related points 
were calculated with a coarser $\mathbf{k}$ mesh
in order to reduce the very large computational cost of the method for these points.
} and the renormalized phonon dispersion curves were obtained by Fourier interpolation.

The strong anharmonicity in this system stabilizes all the imaginary phonon branches even at 0 K,
as can be seen in Fig. \ref{fig1}.  This is an extraordinary effect considering that, normally, 
anharmonic stabilization of unstable modes occurs with increasing 
temperature~\cite{PhysRevLett.100.095901}. 
MD simulations have suggested that sc Ca may be stable
at room temperature~\cite{PhysRevLett.105.235503,PhysRevLett.103.055503}. On the other hand, 
MD calculations cannot~\cite{PhysRevLett.103.055503} predict the stabilization below the
Debye temperature ($\Theta_D\sim 120$ K 
according to our calculations) since, as we demonstrate, sc Ca is stabilized by purely quantum anharmonic effects
at 0 K.
In particular, our results give 26.8 cm$^{-1}$ and 2.6 cm$^{-1}$ for the transverse modes, unstable in the harmonic case,
at the X and M points respectively. 
 Although low-energy transverse modes
show the largest renormalization, longitudinal modes also suffer a considerable anharmonic 
correction.
As expected, phonon frequencies increase when temperature is raised. 
Concretely,  
the temperature dependence is very strong for the transverse mode at M and at R. At 300 K, above $\Theta_D$,
our results are 
in close agreement with the values obtained from MD at the zone boundary 
by Teweldeberham \etal{}~\cite{PhysRevLett.105.235503}.
Note that their calculation was performed at 45 GPa and ours at 50 GPa.
\begin{figure}[t]
\includegraphics[width=0.49\textwidth]{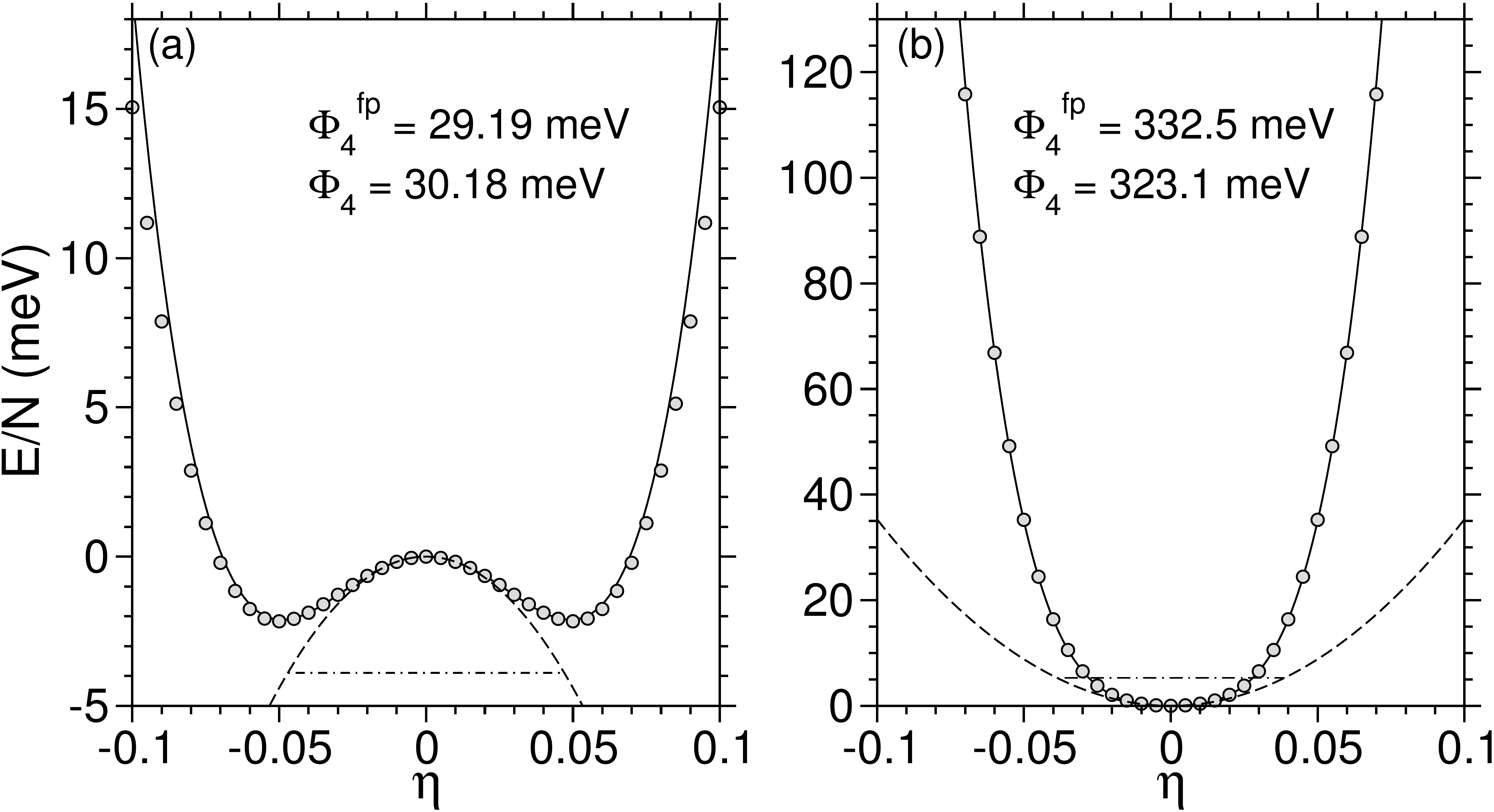}
\caption{Total energy per atom when the atoms are displaced along the transverse mode at M (a) 
and R (b).
The dots depict the \abinitio{} total energies, the solid line is the fit to the quartic potential, 
the dashed line the 
harmonic contribution and the dash-dotted line $\frac{1}{2}\hbar\omega$.
The $\Phi_{4,\nu \mathbf{q}}$ values obtained from the fit to the quartic potential, frozen phonon (fp) result,
and from the differentiation of the dynamical matrices in supercells are shown.}
\label{fig2}
\end{figure}

In sc Ca, as in many other cases~\cite{PhysRevB.68.220509,dastuto:174508,PhysRevB.82.104504}, 
it is crucial to account for 
scattering between phonons with different momenta. Indeed, if we make
the assumption that nondiagonal coefficients are equal to the diagonal ones in Eq. \eqref{gnq}, 
$\Phi^{\alpha_1 \alpha_2 \alpha_3 \alpha_4}(\mathbf{q}',-\mathbf{q}',\mathbf{q},-\mathbf{q})
\sim  \Phi^{\alpha_1 \alpha_2 \alpha_3 \alpha_4}(\mathbf{q},-\mathbf{q},\mathbf{q},-\mathbf{q}) $, 
the error caused in $\Omega_{\nu \mathbf{q}}$ is quite dramatic and the temperature dependence
becomes unrealistic. For example, the mode at R reaches a frequency of 235.5 cm$^{-1}$ at 0 K and 
already 372.2 cm$^{-1}$ at 100 K, 2.6 times larger than our predicted value. 
At R such a difference is a consequence of the huge value of the diagonal coefficients
in comparison to the nondiagonal ones. As shown in Fig. \ref{fig2}, 
the importance of the diagonal anharmonic coefficients can be calculated from frozen phonon calculations. 
For a mode with momentum $\mathbf{q}$ at the edge of the BZ,
when the atoms are displaced from their equilibrium position $\mathbf{R}$ as $\eta a \cos(\mathbf{q}\mathbf{R})
\boldsymbol{\epsilon}_{\nu \mathbf{q}}$, with $a$ the lattice parameter and $\eta$ a small dimensionless number, 
the potential is given as
$E/N(\eta)=\frac{\eta^2}{2}a^2M\omega^2_{\nu \mathbf{q}} + \frac{\eta^4}{4!}\Phi_{4,\nu \mathbf{q}}$, with
\begin{equation}
\Phi_{4,\nu \mathbf{q}} = a^4 \sum_{ \{\alpha\}} 
\epsilon^{\alpha_1}_{\nu \mathbf{q}}
\epsilon^{\alpha_2}_{\nu -\mathbf{q}}
\epsilon^{\alpha_3}_{\nu \mathbf{q}}
\epsilon^{\alpha_4}_{\nu -\mathbf{q}}
\Phi^{\alpha_1 \alpha_2 \alpha_3 \alpha_4}(\mathbf{q},-\mathbf{q},\mathbf{q},-\mathbf{q}).
\label{phi4}
\end{equation}
A fit to this potential yields the frozen phonon $\Phi_{4,\nu \mathbf{q}}$ coefficient. As can be seen in Fig. \ref{fig2},
our values obtained differentiating dynamical matrices in supercells are in good agreement with frozen phonon estimates. 

Our method yields the whole renormalized phonon spectrum at 0 K and, thus,  we can estimate the superconducting transition 
temperature in sc Ca. The usual electron-phonon vertex is not modified by anharmonicity since the matrix elements of the
gradient of the crystal
potential are independent of the phonon frequencies~\cite{grimvall1981}. Therefore, the electron-phonon coupling 
constant $\lambda$ can be calculated straightforwardly using the electron-phonon matrix elements 
and the renormalized frequencies $\Omega_{\nu \mathbf{q}}$ at 0 K. 
The convergence of the electron-phonon matrix elements 
required a denser  $80\times80\times80$ $\mathbf{k}$ grid. Integrating the Eliashberg function,
$\alpha^2F(\omega)$, we obtain $\lambda = 0.74$ and $\omega_{log}= 53$ K, 
leading to $T_c \simeq 2.1$ K, an estimate obtained from the Allen-Dynes modified McMillan 
equation~\cite{PhysRevB.12.905}(we have used $\mu^* = 0.1$ for the Coulomb pseudopotential). 
As can be seen in Fig. \ref{fig1}, the greatest contributions
to $\lambda$ come from the soft transverse modes which are unstable in the harmonic approximation. 
Indeed, the integrated electron-phonon coupling parameter, $\lambda(\omega)$, 
reaches the value of 0.54 at 50 cm$^{-1}$ (that is,  73 \% of the total value of $\lambda$), so that 
if it were not for these soft modes sc Ca would superconduct only below 0.1 $\mu$K.
The value calculated for $T_c$ at 50 GPa is in close agreement with the experimental 1.2 K value obtained by Okada
\etal{}~\cite{JPSJ.65.1924}
and with the 1.7 K value obtained extrapolating linearly the $T_c$ values measured for sc Ca at 
higher pressure in more recent experiments~\cite{JPSJ.75.083703}.
Finally, despite the strong anharmonicity, the isotope coefficient, $\alpha=-\frac{d \ln T_c}{d \ln M}$, 
is predicted to be 0.45, close to the 0.5 value of a BCS superconductor. 

\begin{figure}[t]
\includegraphics[width=0.49\textwidth]{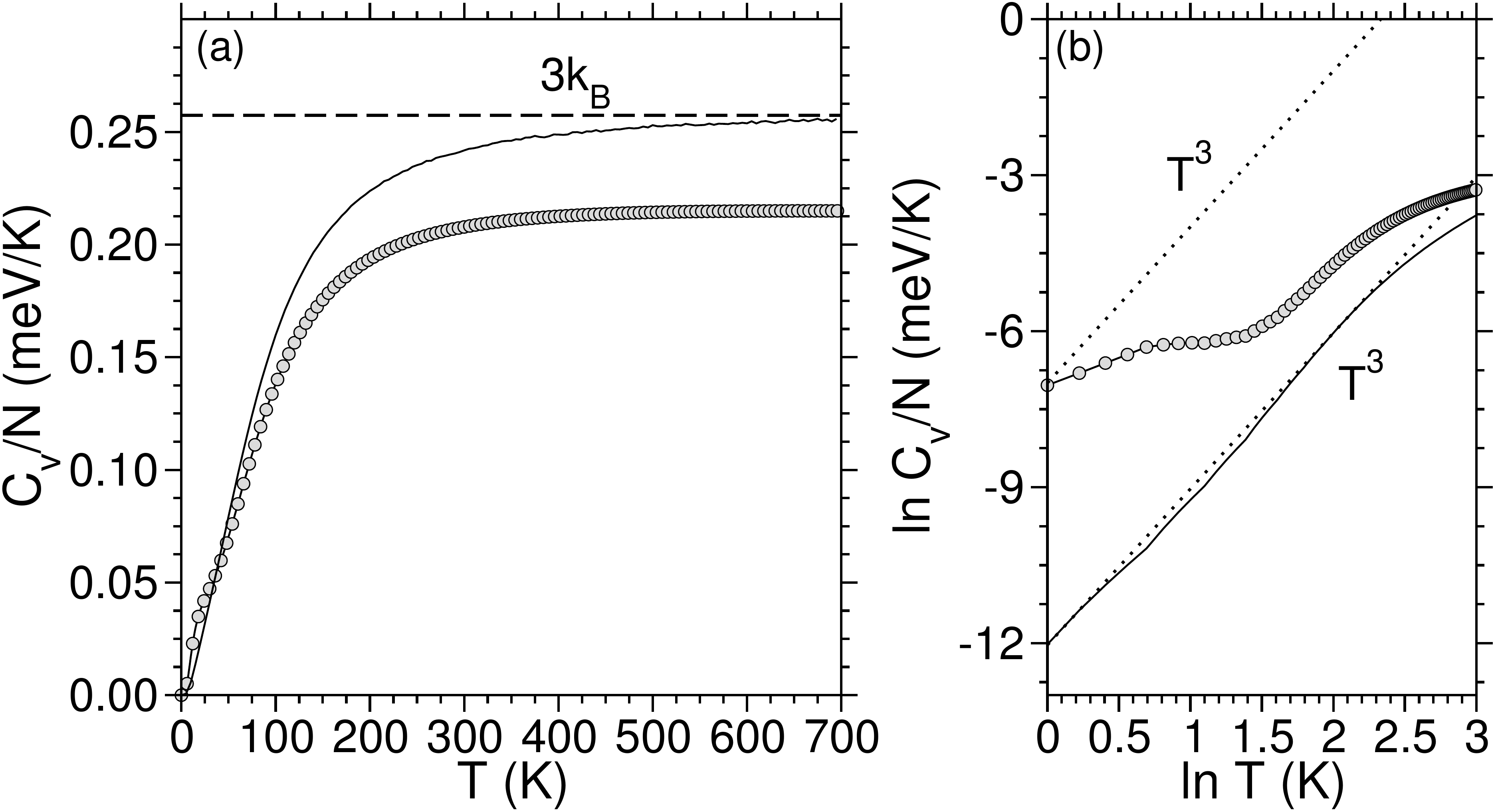}
\caption{(a) Constant volume specific heat per atom for the sc Ca anharmonic crystal (line with grey circles). For comparison,
the specific heat calculated with the
usual harmonic formula including the renormalized $\Omega_{\nu \mathbf{q}}$ frequencies at 0 K is shown (solid line). 
The $3k_B$ line is depicted too (dashed line). 
(b) Low temperature specific heat in logarithmic scale.}
\label{fig3}
\end{figure}

The temperature dependent renormalized frequencies $\{\Omega_{\nu \mathbf{q}}\}$ 
allow us to estimate the anharmonic free energy directly from Eq. \eqref{gibbs-bogoliubov} and the 
constant volume specific heat as 
$C_V=-T\left(\frac{\partial^2 F}{\partial T^2}\right)_V$. As shown in Fig. \ref{fig3}, 
the high-temperature limit of the specific heat per atom is reduced by 17 \% from the classical 
$3k_B$ value given by the equipartition theorem. Such a reduction from the Dulong-Petit law is a sign of
strong anharmonicity~\cite{0022-3719-8-18-012,0022-3719-9-8-011} and has been observed in different 
systems~\cite{PhysRevB.76.014523,PhysRevB.81.134302}. 
Moreover, the low-temperature behavior of $C_V$ is strongly modified from the $T^3$ relation of harmonic crystals.
The anomalies of the specific heat are mainly driven by the temperature dependence of the phonon frequencies in $F_0$. 
Indeed, as depicted in Fig. \ref{fig3}, when the specific heat is calculated from $F_0$ 
assuming that the 0 K renormalized phonons are not modified under temperature, the Dulong-Petit law 
and the low-temperature $T^3$ behavior are recovered as expected. 

In summary, within the SCHA, using a variational procedure based on the Gibbs-Bogoliubov inequality we have shown 
that the high-pressure sc phase of Ca is stabilized even at 0 K by strong quantum anharmonic effects. 
This procedure has been used calculating fully \abinitio{} the anharmonic coefficients up to fourth order in the whole BZ
and may be applied as well in many cases where the phonons are imaginary or anharmonicity needs to be treated beyond
standard perturbation theory. Although below 30 K the sc phase may
transform to a rather similar monoclinic $Cmmm$ phase~\cite{Mao01062010}, which is mechanically unstable in the harmonic
approximation as well and shows very similar harmonic phonons~\cite{PhysRevLett.105.235503}, 
we have calculated the superconducting $T_c$ of sc Ca finding a
 good agreement with experiment. 
Moreover, the huge anharmonicity in this system makes the specific heat very 
anomalous with a large reduction from the high-temperature $3k_B$ limit. An experimental confirmation
of this last feature would indirectly show the strong anharmonic behavior predicted.

The authors are grateful to F. Mauri, M. Calandra and A. Eiguren 
for fruitful discussions and
the Department of Education, Universities and Research of the Basque
Government, UPV/EHU (Grant No. IT-366-07) and the Spanish Ministry of 
Science and Innovation (Grant No. FIS2010-19609-C02-00) for financial support.
Computer facilities 
were provided by the DIPC.

%

\end{document}